# Feasibility of a Small, Rapid Optical/IR Response, Next Generation Gamma-Ray Burst Mission

B. Grossan[1,2], G.F. Smoot[1,4], V.V. Bogomolov[1], S. I. Svertilov[1], N. N. Vedenkin[1], M. Panasyuk[1], B. Goncharov[3], G. Rozhkov[3], K. Saleev[3], E. Grobovskoj[1], A.S. Krasnov[1], V. S. Morozenko[1], V. I. Osedlo[1], E. Rogkov[1], T. V. Vachenko[1], E. V. Linder[1,2]

**Abstract.** We present motivations for and study feasibility of a small, rapid-optical/IR response gamma-ray burst (GRB) space observatory. By analyzing existing GRB data, we give realistic detection rates for X-ray and optical/IR instruments of modest size under actual flight conditions. Given new capabilities of fast optical/IR response (~ 1 s to target) and simultaneous multi-band imaging, such an observatory can have a reasonable event rate, likely leading to new science. Requiring a *Swift*-like orbit, duty cycle, and observing constraints, a *Swift*-BAT scaled down to 190 cm$^2$ of detector area would still detect and locate about 27 GRB yr$^{-1}$ for a trigger threshold of 6.5 σ. About 23% of X–ray located GRB would be detected optically for a 10 cm diameter instrument (~ 6 yr$^{-1}$ for the 6.5 σ X-ray trigger).

## 1. Introduction.

*Swift* has been spectacularly productive in the study of gamma-ray bursts (GRBs), but is past its design lifetime. A new GRB observatory with new capabilities would be welcome. No obvious replacement is on the horizon: the SVOM mission is now uncertain, and other upcoming observatories described as "GRB-capable" lack: (i) high GRB rates, (ii) an on-board optical instrument, and (iii) optical-quality positions. Without (i), you cannot do statistical studies; without (ii) & (iii) you cannot apply ever-evolving techniques in follow-up observations that make *Swift* so productive. Here we study a small post-*Swift* GRB observatory with the following requirements: small, due to limited resources in the current world economy; high GRB rate, to enable statistical studies; new capabilities, to investigate new science; optical quality locations, to enable the most varied possible follow-up science. Is this feasible? How small could a GRB observatory instrument really be?

## 2. The GRB Rate for a "Mini-BAT"

Consider a scaled-down version of *Swift*-BAT for GRB location. From Beppo-Sax to BAT, coded mask X-ray cameras have yielded high GRB rates & localizations smaller than optical telescope fields. In Burrows et al. (2012), the wide field of view of coded mask cameras dominated energy range & sensitivity for maximum GRB rate. Scaling BAT by collecting area implies a high detection rate: SNR (signal to noise ratio) ~ A$^{1/2}$ for steady sources, i.e. rate is *weakly* dependent on A. GRBs are transient, however; their detected light curve, fluence & duration change with background noise, therefore area, so the SNR relation can be more complex. We therefore used actual BAT light curves to predict performance of a smaller BAT, to determine how small an instrument would still trigger & locate GRBs at a high rate.

[1] Extreme Universe Laboratory, Moscow State University, Russian Federation
[2] University of California at Berkeley Space Sciences Laboratory, USA
[3] Moscow State University, Russian Federation
[4] University of California at Berkeley, USA

## 2.1 Method: Rate Estimate from BAT Data $SNR_{peak}$ measurements.

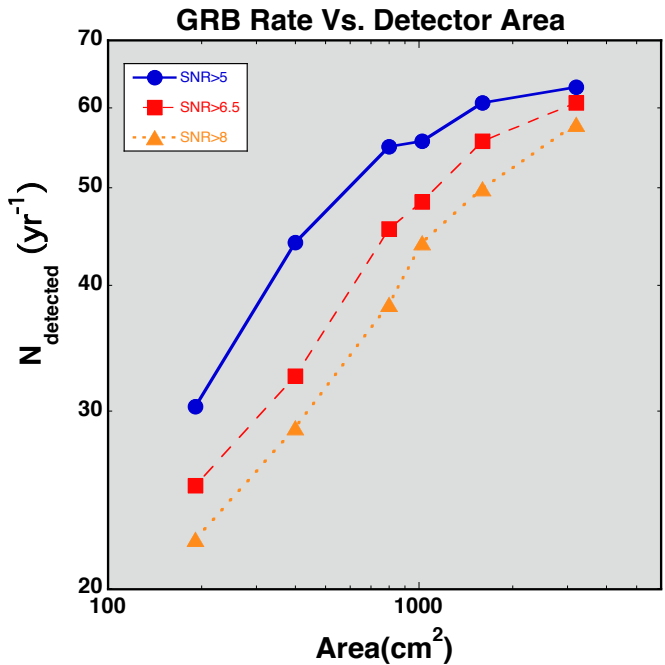

**Fig. 1.** GRB Detection Rate vs. Detector Area for *Swift* BAT-Like Instruments.

The peak SNR time segment of a GRB light curve determines the smallest instrument collecting area, $A_{collect}$, required for its detection. GRB rate as a function of $A_{collect}$ was determined by (i) finding the peak SNR segment in BAT light cuves, (ii) scaling to get $SNR_{peak}(A_{collect})$, then (iii) counting the number of bursts with $SNR_{peak}>$ threshold to determine the detection rate.

We used a very simple $SNR_{peak}$ "Trigger", specifed as follows: We used the sum of 64 ms data channels 1-3 (15-100 kev, the highest S/N combination). Integration time windows of 0.25, 0.5, 1, 2, 4, 8 s were examined for fluctuations > threshold (in σ) over background (the trigger and detection criteria). The trailing average background (t–19.2 to t–6.4 s) was used. All triggers were checked by eye for false triggers (only 1 found). (BAT also has long-window image data trigger, rarely triggered; we had no such trigger, as the benefit for a small instrument would likely be very small.) We analyzed 94 GRB light curves 2010 Nov. - 2012 Mar. to find the $SNR_{peak}$ in each window. We then scaled the SNR for instruments of smaller collecting area, and reported the number of bursts over trigger threshold in the smaller instruments.

Pre-selecting only burst data, as we did, begs the question of false alarm rate. This pre-selection is acceptable because in a real mision, known tools are available to control excessive false alarms: trigger parameter tuning, cutoff rigidity maps, and others.

## 2.2 Analysis Results
Our simplified trigger detected 91% of BAT bursts (86 detections, 1 fail, 7 image trigger non-detects). For only 190 cm$^2$ of collecting area, < 1/25th of *Swift*, 22 GRB/yr would still have $SNR_{trig} > 8$; this number increases to 25 and 30 for $SNR_{trig}$ of 6.5 and 5.0, respectively ([1]).

**Imaging/Location Consistent with Triggering Analysis** After triggering, an image is made, with location uncertainty ~ $1/SNR_{image}$. Is the correlation noise of coded-mask imaging, the dominant noise in $SNR_{image}$, a problem? In a simulation by Connell (2012), all triggers with $SNR_{peak}>5$ yielded $SNR_{image}>8$, the typical coded mask design threshold. Localization quality is therefore not a problem.

**Robust Result** We recognize this approach is valid only for BAT-like instruments with similar orbit, operations & observing constraints. Background depends on instrument & spacecraft construction, via activation & secondary emission. The BAT background is ~ 1.9 cts s$^{-1}$ cm$^{-2}$ 15-150 keV (bat_desc.html). The ESA MXGS coded mask camera, with similar CZT detectors (5 mm thick) and shielding, on the International Space Station (copious mass & solid angle for secondaries), has estimated background = 2.1 cts s$^{-1}$ cm$^{-2}$ 15-200 keV (Renzi). The results are therefore not sensitive to spacecraft platform; orbital inclination and altitude dominate the background considerations.

---

[1] The Ultra-Fast Flash Observatory-pathfinder X-ray camera (UFFO-p; Kim 2012) has 190 cm$^2$ collecting area, but is planned to fly in a polar orbit (89 deg. inclination) with high background regions, losing substantial useful observing time. We roughly estimate a duty cycle of 20% of that of BAT (from the time in high background regions and a 1000 s background decay after). From this, and a field of view 84% that of BAT's, we find that UFFO-p's X-ray rates are ~0.17 of those above, (5.8 yr$^{-1}$, $SNR_{trig}$ =6.5) assuming all else identical to BAT. We then expect ~ 1 detection yr$^{-1}$ for the 10 cm UVOT-like optical instrument on UFFO-p (see sec. 3.1).

### 3. New Science from Follow-Up Optical/IR

A "mini-BAT" would sample the brightest of the underline{known} *Swift* GRB population. How then, do we get new science? *Swift* optical follow-up is hardware-limited to > 60 s after trigger. Using beam-steering mirrors, telescope pointing has been demonstrated in ~ 1 s over similar sized fields for the upcoming Ultra-fast Flash Observatory-Pathfinder (Jeong+). Such rapid-response on-orbit optical/IR follow-up would take advantage of the shorter communication time and lack of weather compared to ground-based rapid follow-up. Such an instrument would yield new information: optical or IR bulk Lorentz factors from the time of opt/IR peak (Molinari+); a much better sample of GRB optical rise times (these rise times are often less than the >60 s *Swift* UVOT response, and so are infrequently measured); the first dust evaporation detection via time-resolved simultaneous multi-band colors (all of which occurs in < 60 s, too fast for most telescopes; see Grossan et al. 2012 for details & additional science topics).

We did not consider a focused X-ray telescope (XRT) for a small observatory, as such instruments are large, expensive, & complex. An IR/optical imager would still yield a precise position for its detections, partially replacing an XRT.

### 3.1 Aperture Size vs. Rate

The brightest *Swift* UVOT V fluxes for each GRB (Fig. 2) show a detection rate strongly dependent on sensitivity. Sensitivity, dominated by non-instrument background, scales as 1/aperture diameter, D (for the same exposure times). For typical conditions, rate decreases by a factor of 0.8 for 2X smaller D (blue line); by 0.7 for a 3X smaller D (pink line). For a 190 (or 1000) cm$^2$ mini-BAT, *Swift* orbit & constraints, the GRB Rate plot (Fig. 1), gives 27 (or 48) GRB X-ray locations/yr (6.5σ trigger). Conservatively neglecting any correlation of V vs. X, for $D_{UVOT}/2$, ~ 7 (or 13) optical detections/yr. are predicted. However, these rates can be significantly increased. Earlier optical acquisition *may* catch many bursts when brighter. A near-IR detector *will* get up to 50% more extinguished GRB (Perley et al. 2009).

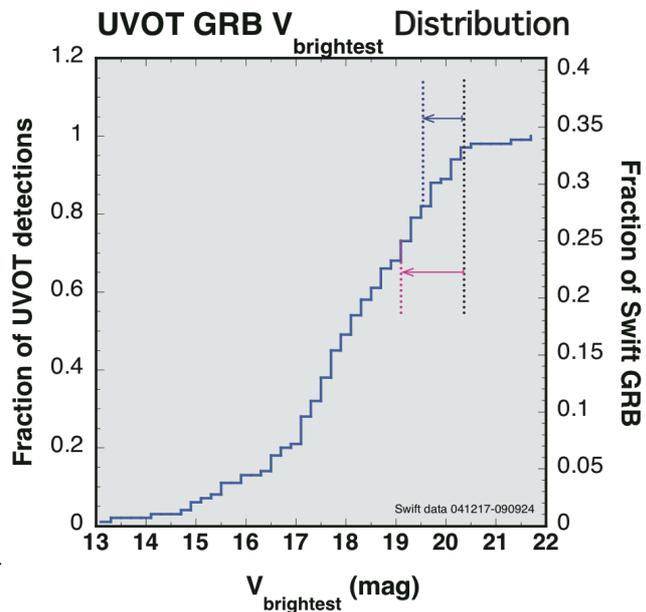

Fig. 2. *Swift* UVOT Maxim Brightness vs. Rate

Detector quantum efficiency (QE) improvement will also boost rate (a CCD has ~ 4-5X UVOT's QE).

### 4. Discussion & Summary

Our conservative analysis using *Swift* data firmly supports the feasibility of small GRB missions: small instruments with good orbits and high duty cycle can produce GRB locations at useful rates for follow-up studies. Improving on *Swift* technology can boost these rates by improving detector sensitivity: e.g., SVOM-like 5-150 keV detectors give ~2.7X BAT (15-150 keV) source photon flux. Better optical QE and IR sensitivity would increase optical/IR rates.

Rapid acquisition for prompt optical emission enables new lines of inquiry. An additional IR channel would yield the first prompt IR measurements and permit the study of dynamic extinction. Additional bands would give more information for small mass cost.

Smaller X-ray cameras can roughly measure GRB durations & spectra, but will have poor Short GRB rates, and greater uncertainty in $E_{peak}$ and flux. We find that new science, even limited to the brightest GRBs of the already known *Swift* population, outweighs these disadvantages. In the future, we will extend this work to investigate the performance of more types of X-ray, optical, IR detectors, spacecraft, and orbits.

**Acknowledgements.** We gratefully acknowledge the support of the Russian Federation Ministry of Education and Science, Agreement No. 11.G34.31.0076.